# Onset of Identical Synchronization in the Spatial Evolution of Optical Power in a Waveguide Coupler

Jyoti Prasad Deka

*Department of Physics, Girijananda Chowdhury University, Guwahati-781017, Assam, India*
*E-mail address: jyotiprasad_physics@gcuniversity.ac.in, jyoti.deka@alumni.iitg.ac.in*

**Abstract:** In this work, we investigated the spatial evolution of optical power in a closed-form optical waveguide configuration consisting of six passive waveguides and each of the waveguides exhibits equal strength of Kerr nonlinearity. We considered only nearest neighbor interaction between the waveguides. We found that in the case of low Kerr nonlinearity, spatial evolution of optical power shows identical synchronization. But when we increased the strength of Kerr nonlinearity, we discovered that spatial evolution of optical power in all waveguides exhibits independent characteristics. On the other hand, we have studied the impact of the coupling constant on the synchronization dynamics of our system. Our findings showed us that strong coupling can strengthen the collective dynamics in the presence of strong Kerr nonlinearity. From our results, we can conclude that Kerr nonlinearity in our system plays the role of disorder parameter that destroys as well as alters the synchronization behavior of evolution of optical power in the waveguides and coupling constant, on the other hand, plays the role of an antagonist and restores synchronization in the model.

## 1. Introduction

Synchronization dynamics is one interesting phenomenon in complex systems that gives rise to numerous interesting dynamical behaviors in such systems. One phenomenon that arises from such complex behavior in nonlinear systems is the emergence of chimera states or the simultaneous existence of coherent and incoherent states in such systems. One of the first mathematical models which pertained to the emergence of chimera states was the Kuramoto model of phase oscillators [1-3]. Kuramoto and his colleagues observed such phenomena in the simulation of arrays of limit-cycle oscillators with non-local coupling. Unihemispherical sleep in the human brain is one such instance where the neuronal network in the human brain exists in both coherent and incoherent states

[4]. Furthermore, such phenomenon has been observed in Josephson junction arrays [5], chemical oscillators [1-3], neuronal conditions such as epileptic seizure [6-7] and Parkinson's disease [8], etc.

In this work, we investigated the spatial evolution of optical power in a closed-form optical waveguide configuration. The mathematical model of the system is based on the Discrete Nonlinear Schrodinger Equation [9-12]. This model is famous for the numerical computation of the spatial evolution of optical power in evanescently coupled waveguides, which are also called Optical Oligomers and phenomenon such as gap solitons has been reported in such systems [10]. In addition, transition to chaos has also been reported in such system [13]. In the recent past, we explored the spatial evolution of optical power a class of optical oligomers which follows the condition of parity-time symmetry [12]. Phenomena such as unidirectionality [14], stable dark solitons in dual core waveguides [15], PT-symmetry breaking in a necklace of coupled optical waveguides [16-17], amplitude death [18], extreme events [19] and so on have been reported in such systems.

In our model, we considered evanescently coupled passive waveguides with nearest neighbour coupling and equal strength of Kerr nonlinearity. Using the prescription described in [12], we converted the mathematical model in the polar coordinates. In this system, we employed the Runge-Kutta Fehlberg Method with a step-size of 0.01 for the numerical simulation. This numerical procedure is an adaptive step-size process and hence, the numerical accuracy in the simulation is far more pronounced. For the evaluation of power spectra of the temporal dynamics of the system, we used the *periodogram* function of MATLAB so as to characterize aperiodic dynamics that arises in our model and also, we discussed the means to possibly control the transition from periodic to aperiodic dynamics.

The manuscript is organized as – section 2 is the mathematical modelling of the system, section 3 is the results and discussion, and section 4 is the conclusion.

## 2. Modelling

We are considering a closed form waveguide coupler exhibiting equal strength of Kerr nonlinearity and equal nearest neighbour coupling as presented in Fig. 1. On top of that, we consider each waveguide is only capable of nearest neighbor evanescent interaction. Hence, using the Discrete Nonlinear Schrodinger Equation (DNLSE) [9], we consider the mathematical model given below.

$$i\frac{dE_j}{dz} = C(E_{j-1} + E_{j+1}) + \beta|E_j|^2 E_j$$

Here, $\beta$ is the strength of nonlinearity, and $C$ is the coupling constant. And here, $z$ is the spatial dimension along the direction of propagation in the optical waveguides and in a way, it could be said that it is also a spatiotemporal variable. All parameters are normalized units.

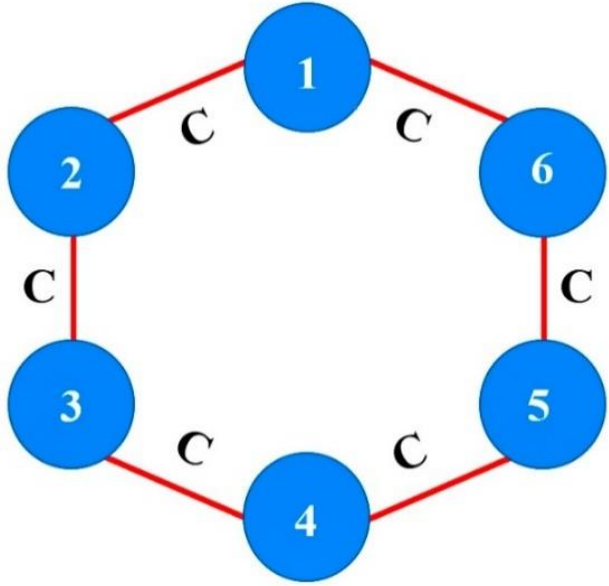

Fig. 1. Schematic of the Coupled Passive Waveguides configuration. $C$ is the coupling strength.

Using $\dot{r}_j = (x_j \dot{x}_j + y_j \dot{y}_j)/r_j$ and $\dot{\theta}_j = (x_j \dot{y}_j - y_j \dot{x}_j)/r_j^2$, this equation could be transformed into a system of nonlinear ordinary differential equations as follows

$$\frac{dr_j}{dz} = C[r_{j-1} \sin(\theta_{j-1} - \theta_j) + r_{j+1} \sin(\theta_{j+1} - \theta_j)] \tag{1a}$$

$$\frac{d\theta_j}{dz} = -\beta r_j^2 - C[r_{j-1} \cos(\theta_{j-1} - \theta_j) + r_{j+1} \cos(\theta_{j+1} - \theta_j)]/r_j^2 \tag{1b}$$

The optical power of the waveguides is given by $P_j = |E_j|^2 = r_j^2$. We then prepare the initial state of the system: $r_j = 1$, $\theta_1 = \theta_2 = \theta_6 = 0$ and $\theta_3 = \theta_4 = \theta_5 = \pi$ and the numerical integrator will have step-size $h = 0.01$. This means that we have launched equal optical power in all waveguides, but half of them have zero-phase and the rest have $\pi$-phase. And this way we could study the influence of phase in the spatial evolution of optical power along the direction of propagation. Moreover, it must be noted that phase lag between two propagation electromagnetic waves is physically equivalent to temporal delay between them. Hence, our model would also enable us to understand how temporal delay could affect the synchronization dynamics of optical power in a waveguide coupler.

## 3. Results and Discussion

Fig. 2 depicts the spatial evolution of optical in all six waveguides for $C = 1$ and $\beta = 0.1$. We can see from Fig. 2(a) that the oscillations of optical power in the 1st and 4th waveguides are of same frequency, phase, and amplitude. But the remaining four waveguides, in comparison, exhibit different optical power. This means that for low strength of Kerr nonlinearity, waveguides display identical synchronization behavior. In our system, we see that all waveguides display synchronization dynamics in two separate bunches of two and four waveguides. It must be noted that we have initially excited the 1st waveguide with zero-phase and the 4th waveguide with $\pi$-phase. So, in a nutshell, we can say that the phase of initial excitation imparted to these waveguides plays a crucial role in the emergence of synchronization behavior in our waveguide coupler configuration.

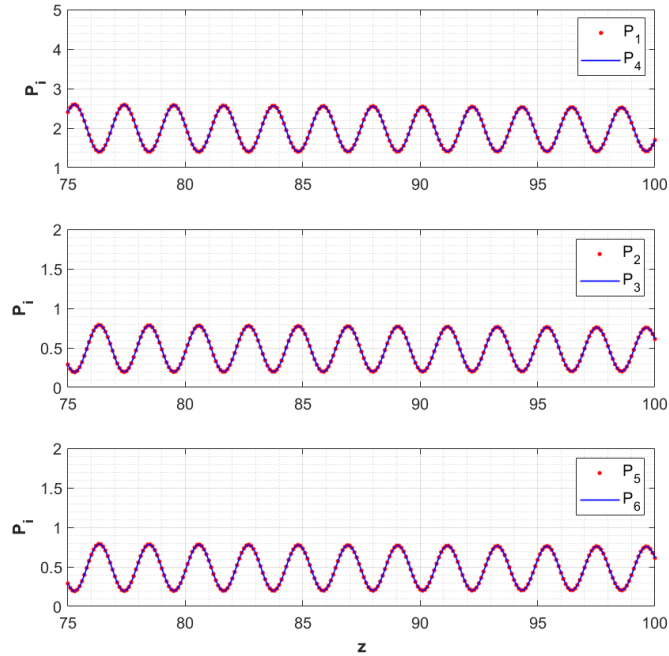

Fig. 2. Spatial Evolution of optical power in (a) 1st, 4th waveguide, (b) 2nd, 3rd waveguide and (c) 5th, 6th waveguide for $C = 1$ and $\beta = 0.1$.

In Fig. 3, we have plotted the spatial evolution of optical power by increasing the strength of Kerr nonlinearity to $\beta = 0.75$. We can see that the spatial evolution shows synchronization dynamics up to $z > 70$ and beyond that, all waveguides start exhibiting aperiodic evolutionary dynamics. On the other hand, it could be seen that now waveguide '**3**' and '**5**' and waveguide '**2**' and '**6**' are now identically synchronized in Fig. 4, where we have plotted the optical power propagating in **3rd** and **5th** waveguides and **2nd** and **6th** waveguides. This is one methodology of depicting the onset of synchronization in coupled systems. When the graph is a straight line and it makes an angle of $\pi/4$ with the x-axis, we say that the two systems are identically synchronized. But on the other hand, when the same makes an angle of $3\pi/4$ with the x-axis, we say it is anti-phase synchronized [20]. From this, we could infer that Kerr nonlinearity plays the role of a disorder parameter which can facilitate desynchronization in our waveguide coupler and also alter the waveguide pair that are identically synchronized.

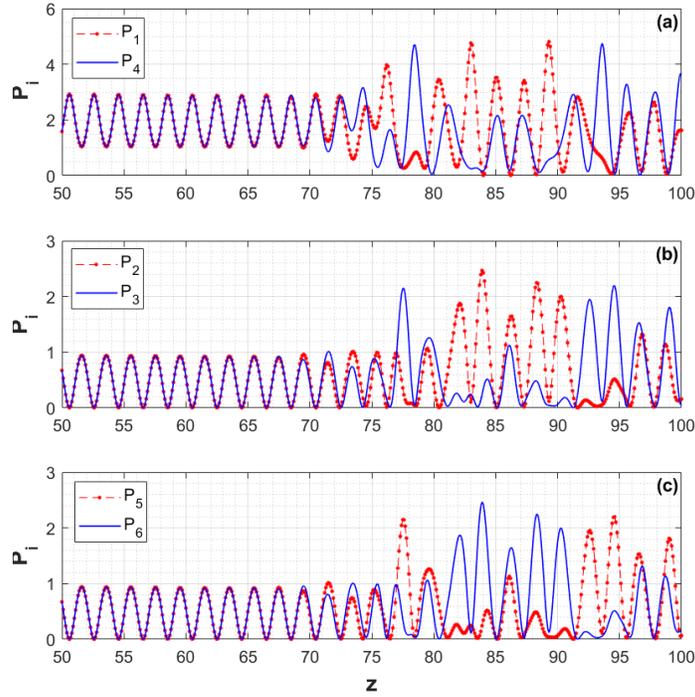

Fig. 3. Spatial Evolution of optical power in (a) $1^{st}$, $4^{th}$ waveguide, (b) $2^{nd}$, $3^{rd}$ waveguide and (c) $5^{th}$, $6^{th}$ waveguide for $C = 1$ and $\beta = 0.75$.

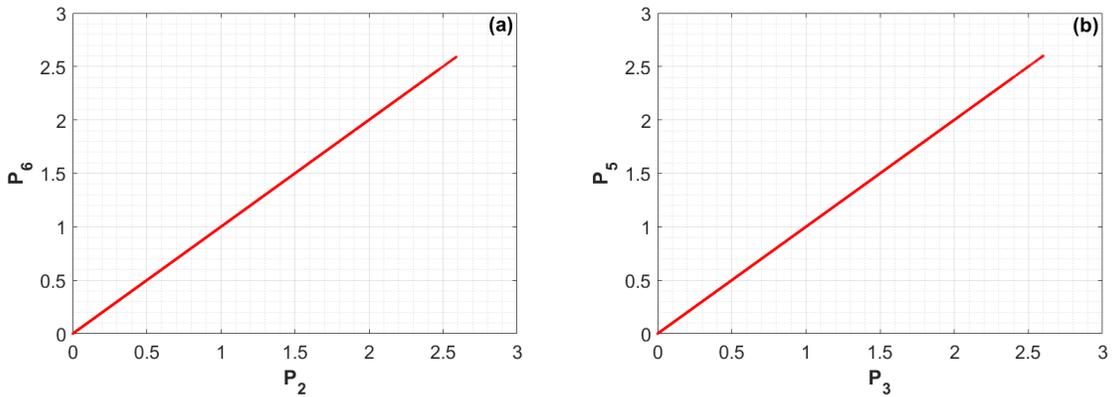

Fig. 4. (a) $P_6$ vs $P_2$ and (b) $P_5$ vs $P_3$ for $\beta = 0.75$ and $C = 1.0$.

In Fig. 5, we have plotted the power spectral density (PSD) of the time series of optical power in the $1^{st}$ waveguide $P_1$ for two cases of the strength of Kerr nonlinearity. We can see from Fig. 5(a) that for low strength of nonlinearity, the power spectrum displays one distinct sharp peak at $f \approx 0.1175$. Here, $f$ is the normalized frequency of the spatial evolution of the system. This is why we observed periodic dynamics in Fig. 2. But when we increase $\beta$, we can see that the power spectrum displays an aperiodic time series in Fig. 5(b) with no distinct peak in the power spectra and in fact, we have observed that in Fig. 3. From this, we can infer that Kerr nonlinearity in our systems is responsible for the transition from periodic to aperiodic dynamics. But when we increased the coupling strength to $C = 3.0$ keeping $\beta = 0.75$ constant, we can see from Fig. 6(a) and 6(b) that the aperiodicity in the spatial evolution of the optical power could be controlled. Now, the power spectrum shows a distinct sharp peak at $f \approx 0.352$ and the spatial evolution is periodic in nature. Furthermore, it should be noted that the power spectrum now depicts a peak at $f \approx 0.352$ and this means that the frequency of oscillation has also increased, and the amplitude too has increased (as observed in Fig. 6(a)). Thus, we can infer that Kerr nonlinearity could also alter the waveguide pair that are identically synchronized and on the other hand, the coupling strength serves as the antagonist agent in the emergence of aperiodic dynamics in our system and also alter the amplitude of oscillating optical power in optical waveguides.

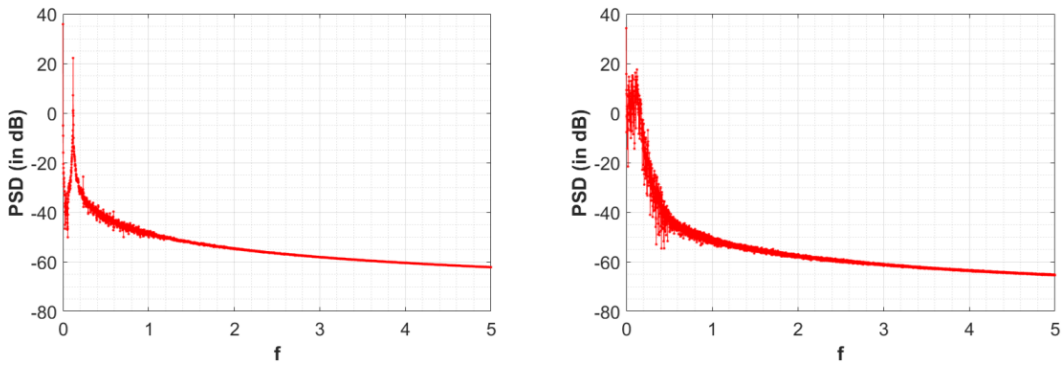

Fig. 5. Power Spectral Density (PSD) of optical power in the 1st waveguide $P_1$ for (a) $\beta = 0.1$ and (b) $\beta = 0.75$.

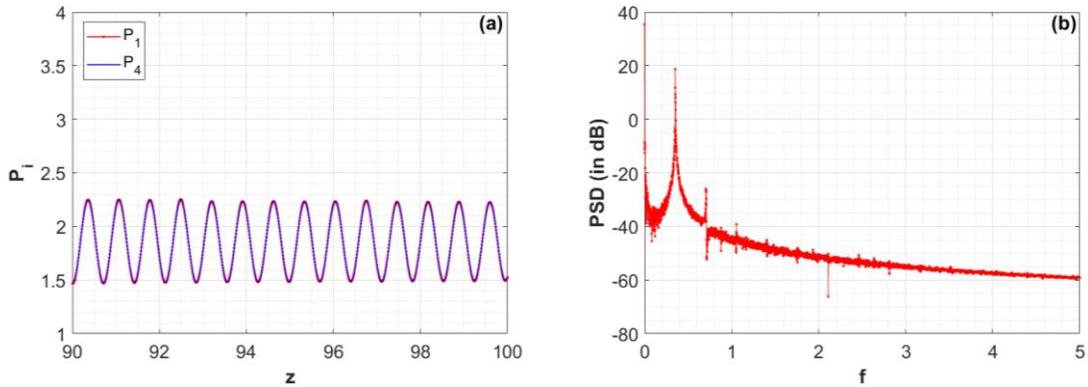

Fig. 6. (a) Spatial Evolution of optical power in the 1st and 4th waveguide and (b) Power Spectral Density (PSD) and for $\beta = 0.75$ and $C = 3.0$.

## 4. Conclusion

In conclusion, we have analyzed the spatial evolution of optical power in a waveguide coupler using the DNLSE as the mathematical model. When the simulation of the evolution of optical power in the system is run considering low strength of Kerr nonlinearity, the optical power in the waveguides is found to be identically synchronized. But when we increase the strength of Kerr nonlinearity but the coupling between the waveguides is kept constant, identical synchronization is observed to a certain limit of the spatial dimension. But beyond that, identical synchronization in the waveguide pairs is destroyed. But the waveguide pairs that are identically synchronized now are changed. So, in short, it could be said that the strength of Kerr nonlinearity could be used as a parameter to initiate synchronization dynamics as well as aperiodic dynamics in the optical power of the waveguides and the same could be used to alter the pair of waveguides that are identically synchronized. On the other hand, to restore identical synchronization between the pair of waveguides that are originally synchronized, one method that we found in our simulation is by increasing the coupling strength between the waveguides. So, in a nutshell, it could be conclusively said that by increasing the strength of Kerr nonlinearity, we can disrupt identical synchronization observed in evanescently coupled waveguides and for the restoration of synchronization, the coupling strength between the waveguides acts as an antagonistic measure.